\documentclass[letterpaper, 10 pt, conference]{ieeeconf}  

\IEEEoverridecommandlockouts                           
\usepackage{amssymb,amsmath,amsfonts,amsthm}
\usepackage{bbm}
\usepackage{bm}
\usepackage{balance}
\usepackage{cite}
\usepackage[T1]{fontenc}
\usepackage[utf8]{inputenc}
\usepackage{graphicx}
\usepackage{soul} 
\usepackage[usenames,dvipsnames]{xcolor}
\usepackage{verbatim}
\usepackage{xspace}
\usepackage{cases}
\usepackage{tikz}
\usepackage[ruled,vlined,linesnumbered]{algorithm2e}
\usepackage{mathtools}

\def\tr{\mathop{\mathrm{tr}}}

\def\CoIL{\mathop{\mathrm{CoIL}}}
\def\VoI{\mathop{\mathrm{VoI}}}
\def\ID{\mathop{\mathrm{ID}}}

\newcommand{\E}{\ensuremath{\mathbb{E}}}
 
\newcommand{\PR}{\mathbb{P}} 
\newcommand{\overbar}[1]{\mkern 1.5mu\overline{\mkern-1.5mu#1\mkern-1.5mu}\mkern 1.5mu}
\newcommand{\norm}[1]{\left\lVert#1\right\rVert}

\DeclareMathOperator*{\argmax}{arg\,max}

\newenvironment{list4}{
	\begin{list}{$\bullet$}{%
			\setlength{\itemsep}{0.05cm}
			\setlength{\labelsep}{0.2cm}
			\setlength{\labelwidth}{0.3cm}
			\setlength{\parsep}{0in} 
			\setlength{\parskip}{0in}
			\setlength{\topsep}{0in} 
			\setlength{\partopsep}{0in}
			\setlength{\leftmargin}{0.2in}}}
	{\end{list}}

\newtheorem{remark}{\bfseries Remark}

\newtheorem{conjecture}{\bfseries Conjecture}

\newtheorem{proposition}{\bfseries Proposition}

\newtheorem{problem}{\bfseries Problem}

\begin{document}
\title{\fontsize{19.7}{23}\selectfont Effect of Computational Power of Sensors on Event-Triggered Control Mechanisms over a Shared Contention-Based Network}

\author{Tahmoores Farjam and Themistoklis Charalambous
	\thanks{T. Farjam and T. Charalambous are with the Department of Electrical Engineering and Automation, School of Electrical Engineering, Aalto University, Espoo, Finland. E-mails: {\tt \{name.surname@aalto.fi\}} }
}

\maketitle

%
%
%
%
\begin{abstract}
In this paper, we study distributed channel triggering mechanisms for wireless networked control systems (WNCSs) for conventional and smart sensors, i.e., sensors without and with computational power, respectively. We first consider the case of conventional sensors in which the state estimate is performed based on the intermittent raw measurements received from the sensor and we show that the priority measure is associated with the statistical properties of the observations, as it is the case of the cost of information loss (CoIL) \cite{Charalambous:2017}. Next, we consider the case of smart sensors and despite the fact that CoIL can also be deployed, we deduce that it is more beneficial to use the available measurements and we propose a function of the value of information (VoI) \cite{Molin:2015,Molin:2019} that also incorporates the channel conditions as the priority measure. The different scenarios and priority measures are discussed and compared for simple scenarios via simulations.
\end{abstract}


%
%
%
%
\section{Introduction}\label{sec:intro}

Modern control environments, such as industrial automation, consist of a multitude of spatially distributed components that are required to exchange information over a shared network. The tremendous increase of data traffic often renders the current transmission protocols incapable to accommodate the required traffic volume. Typically, the capacity constraints of the communication resources are such that the network can only accommodate transmissions only from a limited number of components at any given time. However, in order to meet the performance requirements, methods to share the available communication resources efficiently with respect to a control objective are necessitated. 

For the case of multiple sensors sharing ideal communication channels, it has been proved that transmissions based on a periodic schedule is optimal \cite{Mo:2014}. Consequently, the static optimal transmission sequence can be computed offline and distributed channel access can be provided by using time division multiple access (TDMA). Finding the optimal solution to time-based sensor scheduling problems in various scenarios has been an active area of research; see, for example, \cite{Park:2018} and references therein. Finding the optimal solution is subject to the curse of dimensionality and thus such methods are often inapplicable in large-scale networks. 

To circumvent this, sub-optimal yet efficient contention-based dynamic scheduling methods can be utilized to allocate the resources based on time-varying transmission priorities. Try-once-discard (TOD) is one of the most well-known such methods which prioritizes transmission according to the deviation of the state from its nominal value \cite{Walsh:2001a}. Although originally proposed for deterministic systems and ideal channels, its application has been extended to stochastic systems with full state observation \cite{Mamduhi:2018} as well as partial observations \cite{Molin:2015, Molin:2019} and non-ideal communication channels \cite{Christmann:2014}. Instead of using measurement-based priorities and bit-wise contention resolution as the aforementioned works, a timer-based mechanism was proposed in \cite{Farjam:2018} which prioritizes transmissions based on a variance-based measure for reducing communication overhead and its application was later extended to wireless networks with uncorrelated \cite{Farjam:2019a,Farjam:2021} and correlated packet dropouts \cite{Farjam:2019b}.

In a separate strand of research, adopting carrier-sense multiple access with collision avoidance (CSMA/CA) protocol and utilizing event-triggered methods, where data transmission is triggered only upon occurrence of certain events has been extensively studied; see \cite{Ge:2020}. Such methods offer easier implementation and lower computational complexities compared to optimal periodic schedules. Furthermore, they can lead to lower communication frequency compared to timer-triggered approaches which in turn results in less congestion on the network and thus less packet dropouts. Hence, a common approach has been to design the triggering rules for each subsystem independently based on the assumption that packet transmissions are always successful \cite{Wu:2013, Han:2015, Trimpe:2014, Trimpe:2015}. However, it was shown in \cite{Xia:2017} that performance of such methods can deteriorate the performance of the system significantly when multiple subsystems compete for the available resource in practical scenarios. In addition, collision-free transmission can also lead to packet dropouts when using wireless communication, which has rarely been addressed except recently \cite{Leong:2017a, Leong:2017, M.HadiBalaghi:2018}. However, to the best of our knowledge, the effect of channel access decisions when multiple event-triggered subsystems compete for transmission over the a shared non-ideal (unreliable) network has not been addressed yet.

In this paper, we study the effect of sensors' computational capabilities on the performance of triggering mechanisms when the effect of channel access decisions in a wireless network is taken into consideration. This is inspired by \cite{Trimpe:2014} where, several triggering laws which are compatible with the computation power of scalar sensors monitoring a single process, are considered. In case of conventional sensors variance-based triggering (VBT) is considered which is closely related to the cost of information loss (CoIL) developed for multiple time-triggered vector processes in \cite{Charalambous:2017, Farjam:2018}. Similarly, the measurement-based triggering (MBT) for smart sensors has been considered for time-triggered vector systems in \cite{Molin:2015, Molin:2019, Ayan:2019, Soleymani:2016} under the label value of information (VoI). 
The contributions of this paper are the following.
\begin{list4}
\item We first consider the case of multiple subsystems equipped with conventional sensors that share a wireless network. For a triggering policy designed for energy conservation, we derive the priority measure based on the information available to the decision makers, in this scenario the estimators, which is shown to be a function of CoIL. It is the first time that CoIL is used in an event-triggered control mechanism for a WNCS consisting of multiple subsystems. 
\item Next, the sensing and estimation architecture for smart sensors is considered. While in this case the priority measure can again be a function of CoIL, we adopt the concept of VoI. More specifically, while VoI has been
developed for resource allocation over perfect channels \cite{Molin:2015, Molin:2019}, in this case, we consider VoI for unreliable channels channels and derive its closed form expression. Despite the wide adoption of VoI in the literature, this is the first time the concept of VoI is connected with unreliable communication channels which also allows for distributed implementation.
\item In both cases, the derived priority measures allow for distributed implementation and their practical realization over TOD is proposed. 
\end{list4}

The remainder of the paper is organized as follows. Section~\ref{sec:model} provides the necessary preliminaries and system model. In Section~\ref{sec:coil} and \ref{sec:voi}, we present the sensing and estimation architecture and the corresponding priority measures for the case of conventional and smart sensors, respectively. We demonstrate the performance of the proposed schemes in Section~\ref{sec:results} and finally draw conclusions and discuss future directions in Section~\ref{sec:conclusions}.

\emph{Notation:}
Vectors and matrices are denoted by lowercase and uppercase letters, respectively. A random vector $x$ from a multivariate Gaussian distribution with mean vector $\mu$ and covariance matrix $X$ is denoted by $x \sim \mathcal{N} (\mu, X)$. The Euclidean norm of a vector $x$ is denoted by $\norm{x}$ and $\sigma_{\max}(X)$ denotes the spectral radius of a matrix $X$. $\mathbb{S}_+^n$ is the set of $n$ by $n$ positive semi-definite matrices. The transpose matrix of matrix $A$ is denoted with $A^{T}$ and its inverse with $A^{-1}$. $f^n(\cdot)$ is the $n$-fold composition of $f(\cdot)$ with the convention that $f^0(X)=X$, and $g\circ f (\cdot) \triangleq g(f(\cdot))$. $\mathbb{E}\{\cdot \}$ represents the expectation of its argument. The $n$ by $n$ identity matrix is represented by $I_n$.

%
%
%
%
\section{Problem Formulation}\label{sec:model}
\begin{figure}[t]
	\centering
	\includegraphics[width=.95\columnwidth]{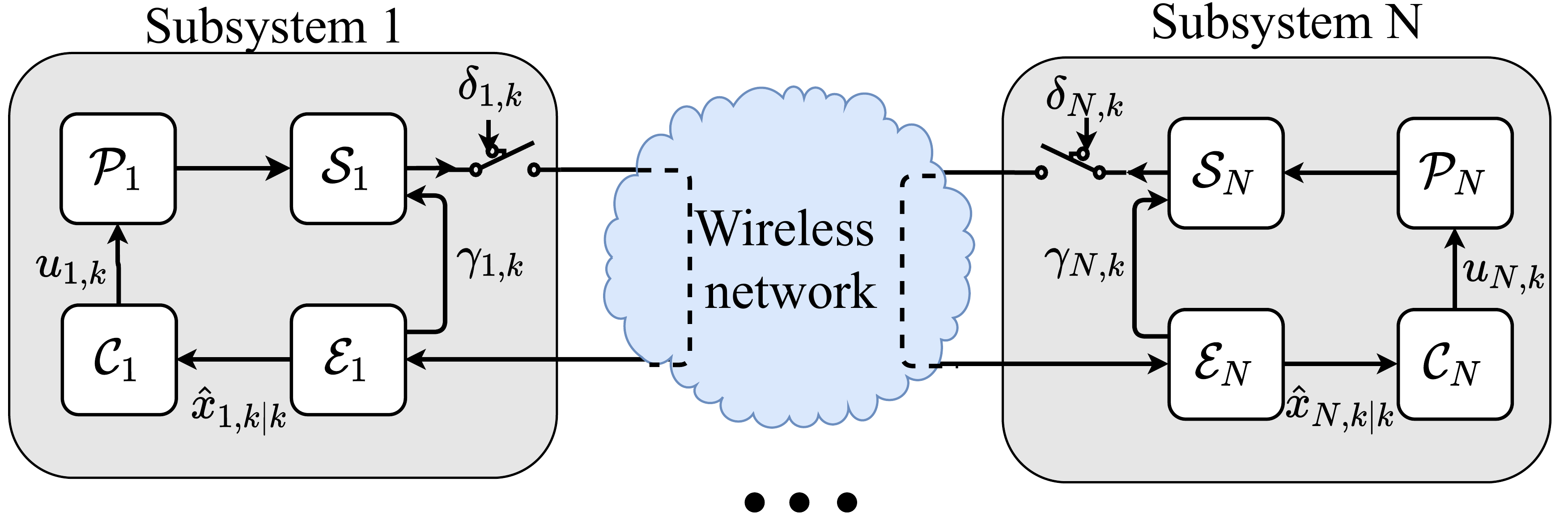}
	\vspace{-0.25cm}
	\caption{Schematic diagram of $N$ subsystems competing for access over a shared wireless network. $\mathcal{P}_i$ represents the plant of subsystem $i\in\{1,\ldots,N\}$, with $\mathcal{S}_i$, $\mathcal{E}_i$, and $\mathcal{C}_i$ being its sensor, estimator and controller, respectively.}
	\label{fig:diagram}
\end{figure}

We consider the scenario in which $N$ dynamical subsystems use a shared communication network to accomplish their control tasks. Each subsystem $i\in\{1,\dots,N\}$ consists of a plant ($\mathcal{P}_i$), dedicated sensor ($\mathcal{S}_i$), estimator ($\mathcal{E}_i$), and controller ($\mathcal{C}_i$); see Fig.~\ref{fig:diagram}. Packet transmission from $\mathcal{S}_i$ to $\mathcal{E}_i$ is supported by a time-slotted shared network. We consider the scenario in which the network consists of a wireless channel (which by definition it is non-ideal), where i.i.d. packet dropouts are possible even when channel access is collision-free. The extension of the proposed scheme to networks with multiple wireless channels is straightforward as it will be discussed in Remark~\ref{remark:2}.
\subsection{Plant and outputs}
The states of each subsystem $i\in\{1,\dots,N\}$ evolves according to the following linear time-invariant (LTI) process: 
\begin{align}
	x_{i,k+1} = A_{i}x_{i,k}+B_{i}u_{i,k}+w_{i,k}, \label{eq:process}
\end{align}
where $x_{i,k} \in \mathbb{R}^{n_{i}}$ and $u_{i,k} \in \mathbb{R}^{m_{i}}$ are the states and inputs at time step $k$, respectively, with $A_{i}$ and $B_{i}$ being the system and input matrices of appropriate dimensions. Furthermore, $w_{i,k}\in \mathbb{R}^{n_{i}}$ is the i.i.d. process disturbance with $w_{i,k}\sim \mathcal{N}(0,W_i)$ and the initial state is $x_{i,0}\sim\mathcal{N} (\bar{x}_{i,0},X_i)$.

The output measured by the sensor is given by
\begin{align}
	y_{i,k} = C_{i}x_{i,k}+v_{i,k}, \label{eq:sensor}
\end{align}
where $C_{i} \in \mathbb{R}^{p_{i}\times n_i}$ is the output matrix and $v_{i,k}\in \mathbb{R}^{p_{i}}$ is the i.i.d. measurement noise described by $v_{i,k}\sim \mathcal{N}(0,V_i)$. We assume that $w_{i,k}$, $v_{i,k}$ and $x_{i,0}$ are mutually independent.
\subsection{Controller and the quadratic cost}
The aim of the controller is to minimize the standard quadratic cost over the infinite horizon which is given by	
\begin{align}\label{eq:Jinfty}
	J_{0:\infty} = \lim\limits_{K \to \infty}\frac{1}{K}\mathbb{E}\left\{\sum_{k=0}^{K-1}\sum_{i=1}^N\left(x^T_{i,k}Q_ix_{i,k}+u^T_{i,k}R_i u_{i,k}\right)\right\},
\end{align}
where $Q_i$ and $R_i$ are weighting matrices of appropriate dimensions. The controller uses the certainty equivalence law and is given by \cite{Molin:2015} 
\begin{align} \label{eq:u}
	u_{i,k} = L_{i,\infty}\hat{x}_{k|k},
\end{align}
where $\hat{x}_{k|k}$ is the \emph{a posteriori} state estimate provided by $\mathcal{E}_i$. Furthermore, $L_{i,\infty}$ is the constant feedback gain which is given by 
\begin{align} \label{eq:L}
	L_{i,\infty} = -(B^T_i\Pi_{i,\infty} B_i + R_i)^{-1}B^T_i\Pi_{i,\infty} A_i,
\end{align}
and $\Pi_{i,\infty}$ is the solution of the following discrete-time algebraic Riccati equation (DARE)
\begin{align*} 
	\Pi_{i,\infty}=A_i^T \Pi_{i,\infty} A_i + Q_i - L_{i,\infty}^T(B_i^T \Pi_{i,\infty} B_i+R_i)L_{i,\infty}.
\end{align*}
By assuming that the pairs $(A_i,B_i)$ and $(A_i,Q_i^{1/2})$ are controllable and observable, respectively, the given DARE has a unique positive semi-definite solution $\Pi_{i,\infty}$ \cite{BrianD.O.Anderson:2012}. By using the proposed controller, the single step quadratic cost at $k$ can be written as \cite{Molin:2015, Charalambous:2017}
\begin{align} \label{eq:Jk}
J_{k} =\sum_{i=1}^NJ_{i,k},
\end{align}
where
\begin{align} \label{eq:Jik}
J_{i,k}=\tr\left(\Pi_{i,\infty} W_{i} \right) + \tr(\Gamma_{i,\infty}\E\{e_{i,k|k}e_{i,k|k}^T\}),
\end{align}
where $e_{i,k|k}\triangleq x_{i,k}-\hat{x}_{i,k|k}$ and $\Gamma_{i,\infty} = L_{i,\infty}^T(B_i^T \Pi_{i,\infty} B_i + R_i)L_{i,\infty}$.

\subsection{Capacity constrained network}
We consider the case where the triggering criterion and associated thresholds are prespecified as required by the available energy budget and let $\theta_{i,k}=\{0,1\}$ denote whether subsystem $i$ competes for transmission at $k$. In case the triggering threshold is crossed $\theta_{i,k}=1$, and $\theta_{i,k}=0$ otherwise. We assume that all the nodes on the network are synchronized and the duration of a transmission frame is less than the sampling time of subsystems and thus the effect of delay can be ignored. The channel access decision is given by
\begin{align}
	\delta_{i,k}=\begin{cases}
		1, &\text{if $\theta_{i,k}=1$ and sensor $i$ transmits at $k$},\\
		0, &\text{otherwise.}
	\end{cases}
\end{align}
To ensure that channel access is collision-free, we impose the following constraint
\begin{align} \label{eq:cnstrnt1}
	\sum_{i=1}^N\delta_{i,k}\leq 1,\quad \forall k\geq 0.
\end{align}
Due to the unreliable nature of the wireless medium, transmitted data packets might not be received successfully at the receiver, i.e., the corresponding estimator. We assume acknowledgement/negative-acknow\-ledge\-ments (ACK/NACK) feedback mechanism is in place which informs the transmitter about the status of the sent packet. This can be represented as another binary variable $\gamma_{i,k}$ which is defined as
\begin{align}
	\gamma_{i,k}=\begin{cases}
		1, &\text{if $\delta_{i,k}{=}1$ and packet is successfully received},\\
		0, &\text{otherwise,}
	\end{cases}
\end{align}
We consider the case of memoryless wireless channels where the packet dropouts are i.i.d. random and the probability of successful transmission over each communication link is given by
\begin{align}
	q_{i}=\PR\{\gamma_{i,k}=1|\delta_{i,k}=1\}.
\end{align}

\subsection{Contention resolution}
In this work, we intend to utilize TOD protocol for collision-free distributed channel access. In the celebrated work \cite{Walsh:2001a}, dynamic identifiers, which depend on the performance criterion, were proposed for contention resolution over wired networks and its application was later extended to wireless networks in \cite{Christmann:2014}. Here, we briefly review how this protocol operates.

\begin{figure}[h]
	\centering
	\includegraphics[width=.95\columnwidth]{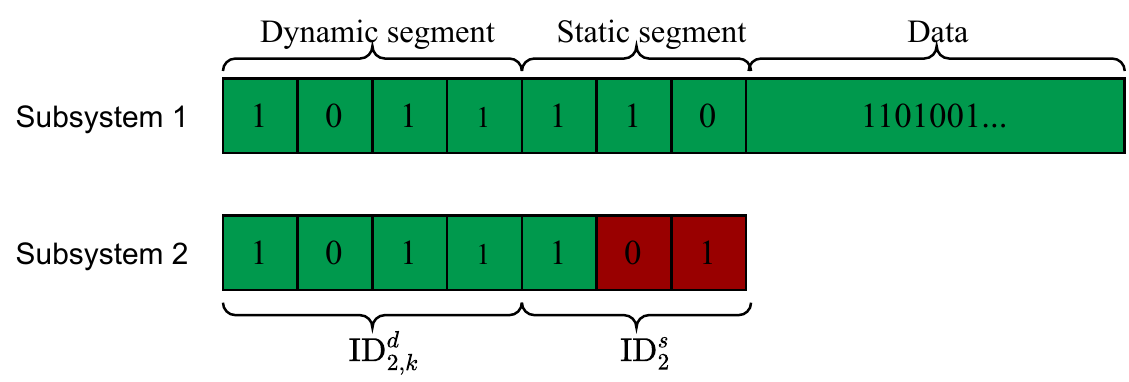}
	\vspace{-0.25cm}
	\caption{Example of two subsystem competing for transmission at $k$. The first $7$ bits are dedicated to contention resolution with $n=4$ bits for the dynamic identifier in \eqref{eq:TOD}. Although ${\ID}_{1,k}^d={\ID}_{2,k}^d$, assuming $1$ is the dominant bit, channel access is granted to Subsystem 1 since it has the dominant static identifier.}
	\label{fig:TOD}
\end{figure}

Let ${\ID}_{i,k}$ denote the identifier of subsystem $i$, at time step $k$, which represents its priority, which consists of a time-varying dynamic segment (${\ID}_{i,k}^d$) and a time-invariant static segment (${\ID}_{i}^s$). In the beginning of each frame, subsystems compete for channel access based on ${\ID}_{i,k}$ and the one with the highest priority, i.e., dominant ${\ID}_{i,k}$, claims the channel. Let $f(\cdot)$ be a continuous, nonnegative, and  monotonically non-decreasing function; then, the dynamic identifier is determined by ${\ID}_{i,k}^d = [f(m_{i,k})]$ where $[\cdot]$ denotes the function round to the nearest integer and $m_{i,k}$ denotes the priority measure. Assuming that the dynamic segment consists of $n$ contention bits, it is constrained between $0$ and ${\ID}_{\text{max}}^d$, where ${\ID}_{\text{max}^d} = 2^n -1$ denotes the identifier's upper bound and thus the dynamic identifier assignment could be defined as
\begin{equation}\label{eq:TOD}
	{\ID}_{i,k}^d=
	\begin{cases}
		0 & \text{if $ {\ID}_{i,k}^d \leq 0$,  }\\
		[f({m}_{i,k})]& \text{if $ {\ID}_{i,k}^d \leq {\ID}_{\text{max}}^d$,}\\
		{\ID}_{\text{max}}^d& \text{otherwise.}
	\end{cases}
\end{equation}
Moreover, ${\ID}_{i}^s$ is a prespecified unique identifier assigned to each subsystem. Although it is possible for multiple subsystems to have the same dynamic identifier in \eqref{eq:TOD}, the unique ${\ID}_{i}^s$ ensures that channel access is provided in a collision-free manner satisfying \eqref{eq:cnstrnt1}. Fig.~\ref{fig:TOD} illustrates how contention is resolved between two subsystems with the same dynamic identifier competing for channel access at $k$.
\subsection{Problem of interest}
Our objective is distributed allocation of the available communication resources such that the following problem is solved.
\begin{problem} \label{problem:p1}
	\begin{align}
		\begin{split}
			\min_{\delta_{i,k}} &\quad \E\{{J}_k|\mathcal{I}^k\}, \\
			\text{subject to}&\quad \eqref{eq:cnstrnt1},
		\end{split}
	\end{align}
where ${J}_k$ is given in \eqref{eq:Jk} and $\mathcal{I}^k$ denotes the information available at the decision makers.
\end{problem}

%
%
%
%
\section{Scenario 1: Conventional Sensors}\label{sec:coil}
In this section, we consider the scenario in which conventional sensors, i.e., sensors with no computational capability, are utilized. First, we discuss how the state estimate is calculated at the estimator side based on the intermittent raw measurements received from the sensor. Then, we prove that the solution of Problem~\ref{problem:p1} can be obtained by utilizing CoIL when the estimators act as the decision makers.

\subsection{Sensing and estimation}
In this scenario, all sensors are assumed to lack any computational power and in the event sensor $i$ receives a transmission request at $k$, i.e., $\delta_{i,k}=1$, it sends the data packet containing its most recent raw measurement, i.e., $y_{i,k}$ in \eqref{eq:sensor}. Consequently, the information available at the corresponding estimator at $k$ is given by $\mathcal{I}_{i,k}=\{ \delta_{i,0},\gamma_{i,0}y_{i,0},\dots,\delta_{i,k},\gamma_{i,k}y_{i,k} \}$. Define the \emph{a priori} and \emph{a posteriori} state estimates and the corresponding error covariances as
\begin{align*}
	\hat{x}_{i,k|k-1}&\triangleq \E\{x_{i,k} | \mathcal{I}_{i,k-1}\},\qquad \hat{x}_{i,k|k}\triangleq \E\{x_{i,k} | \mathcal{I}_{i,k}\},\\
	P_{i,k|k-1}&\triangleq\E\{(x_{i,k}-\hat{x}_{i,k|k-1})(x_{i,k}-\hat{x}_{i,k|k-1})^T | \mathcal{I}_{i,k-1}\},\\
	P_{i,k|k}&\triangleq\E\{(x_{i,k}-\hat{x}_{i,k|k})(x_{i,k}-\hat{x}_{i,k|k})^T | \mathcal{I}_{i,k}\},
\end{align*}
respectively. For simplicity, we assume that the estimator ignores any additional information associated with $\delta_{i,k}=0$. Thee celebrated result of Kalman filtering with intermittent observations can then be applied in this setting \cite{Sinopoli:2004}, where $\hat{x}_{i,k|k}$ fed to the controller is obtained by
\begin{subequations}
	\begin{align}
		\hat{x}_{i,k|k-1} &= A_{i}\hat{x}_{i,k-1|k-1}+B_iu_{i,k-1},\\
		P_{i,{k|k-1}} &= h_i(P_{i,{k-1|k-1}}),  \\
		K_{i,k} &= P_{i,{k|k-1}} C_i^T( C_iP_{i,{k|k-1}} C_i^T + V_i )^{-1},\\
		\hat{x}_{i,k|k} &= \hat{x}_{i,k|k-1} + \gamma_{i,k}K_{i,k} (y_{i,k}-C_i\hat{x}_{i,k|k-1}),  \label{eq:posxE} \\
		P_{i,{k|k}} &=\begin{cases}
			P_{i,{k|k-1}}, & \text{if } \gamma_{i,k}=0,\\
			g_i\circ h_i(P_{i,{k-1|k-1}}), & \text{otherwise},
		\end{cases} \label{eq:PposE}
	\end{align}
\end{subequations}
where the functions $h,\,g: \mathbb{S}_+^n \to \mathbb{S}_+^n$ are defined as
\begin{align}
	h_i(X) &\triangleq A_i X A_i^T + W_i, \label{eq:h}\\
	g_i(X) &\triangleq X - X C_i^T (C_i X C_i^T + V_i)^{-1} C_i X. \label{eq:g}
\end{align}

\subsection{Priority measure and contention resolution}
Due to the lack of computational power at the sensor, basic triggering mechanisms such as \emph{send-on-delta} (SoD) can be implemented at the sensor. However, it has been shown that utilizing the computational power at the estimator side for triggering the transmission can greatly enhance performance despite the lack of actual measurements \cite{Trimpe:2014, Trimpe:2015}. This is also known as \emph{variance-based triggering} (VBT) where once the value of error covariance at the estimator exceeds a specified threshold, it sends a transmission request to the corresponding sensor. Proposition~\ref{prop:p1} shows how CoIL can be utilized in this context for prioritizing channel access. Note that estimators compete for claiming the channel at the start of frame $k$. Thus, the information available to each decision maker is $\mathcal{I}_{i,k-1}$ and $\mathcal{I}^k\triangleq\cup_{i\in\{1,\ldots,N\}}\mathcal{I}_{i,k-1}$ in Problem~\ref{problem:p1}.
\begin{proposition} \label{prop:p1}
	Let $\mathcal{F}_{k}\triangleq\{i:i\in\{1,\dots,N\},\theta_{i,k}=1\}$ denote the set of subsystems which cross the triggering threshold. Define CoIL for subsystem $i$ at $k$ as
	\begin{align} \label{eq:coil}
		{\CoIL}_{i,k}\triangleq\tr\left(\Gamma_{i,\infty}\left[P_{i,k|k-1}{-}g_i\circ h_i(P_{i,{k-1|k-1}})\right]\right).
	\end{align}
	Then Problem~\ref{problem:p1} is solved by letting $\delta_{i^*,k}=1$ where 
	\begin{align} \label{eq:istrCoIL}
		i^*=\argmax_{i\in\mathcal{F}_{k}}  \quad {\CoIL}_{i,k}q_i.
	\end{align}
\end{proposition}
\begin{proof}
By following the approach of \cite[Lemma 2]{Charalambous:2017, Farjam:2021}, Problem~\ref{problem:p1} can be equivalently written as
\begin{align} \label{eq:istrJ}
	i^*=\argmax_{i\in\mathcal{F}_{k}}  \quad& \E\{{J}_{i,k}|\mathcal{I}_{i,k-1},\delta_{i,k}=0\}\notag\\
	&\quad -\E\{{J}_{i,k}|\mathcal{I}_{i,k-1},\delta_{i,k}=1\}.
\end{align}
From \eqref{eq:Jik} and the law of total expectation we obtain
\begin{align} \label{eq:Jik0}
	&\E\{{J}_{i,k}|\mathcal{I}_{i,k-1},\delta_{i,k}=0\}=\tr\left(\Pi_{i,\infty} W_{i} \right) \notag \\ 
	&\qquad \qquad \qquad  +\tr(\Gamma_{i,\infty}\E\{e_{i,k|k}e_{i,k|k}^T|\mathcal{I}_{i,k-1},\delta_{i,k}=0\})\notag \\ 
	& \quad = \tr\left(\Pi_{i,\infty} W_{i} \right)+\tr(\Gamma_{i,\infty}P_{i,k|k-1}),
\end{align}
where the second equality follows from \eqref{eq:PposE}. Similarly, 
\begin{align} \label{eq:Jik1}
	&\E\{{J}_{i,k}|\mathcal{I}_{i,k-1},\delta_{i,k}=1\}=\tr\left(\Pi_{i,\infty} W_{i} \right) \\ 
	& {+} (1{-}q_i)\tr(\Gamma_{i,\infty}P_{i,k|k-1}) {+} q_i\tr(\Gamma_{i,\infty}g_i\circ h_i(P_{i,{k-1|k-1}})).\notag
\end{align}
Finally, substituting \eqref{eq:Jik0} and \eqref{eq:Jik1} in \eqref{eq:istrJ} yields \eqref{eq:istrCoIL}.
\end{proof}
Since computation of ${\CoIL}_{i,k}q_i$ requires no information exchange between subsystems, from Proposition~\ref{prop:p1} it follows that Problem~\ref{problem:p1} can be solved in a distributed fashion. In this regard, TOD can be implemented for this purpose by assigning $m_{i,k}= {\CoIL}_{i,k}q_i$ in \eqref{eq:TOD}. Note that due to the limited number of contention bits $n$, this scheme does not necessarily provide the optimal solution to Problem~\ref{problem:p1}, since multiple subsystems might be assigned with the dominant dynamic identifier at some $k$. Nevertheless, the mapping function $f(\cdot)$ in \eqref{eq:TOD} can be fine-tuned based on the network configuration to increase the probability of the subsystem with the largest value for ${\CoIL}_{i,k}q_i$ having the dominant dynamic identifier. During the contention period, all sensors listen to the shared medium. Once the contention period ends, the sensor corresponding to the dominant estimator can infer permission to transmit from the unique static identifier and sends its data packet without any collisions.

\begin{remark}
	Neither the triggering mechanism nor the contention resolution require the actual measurements from the sensor. Implementing VBT enables the estimator to determine the exact number of sampling times that the triggering condition will not be met after it successfully receives a packet. During this period, the sensor can enter sleeping mode since measurements and listening on the medium is unnecessary. Therefore, energy consumption can be further reduced by sending the sleeping duration to the sensor in addition to ACK/NACK through the feedback channel.
\end{remark}

\begin{remark}\label{remark:2}
	This scheme can be easily extended to networks with multiple channels $j\in\{1,\dots,M\}$. In this scenario, an additional constraint is added, such that each subsystem occupies a single channel only at any specific time slot $k$, i.e., $\sum_{j\in\{1,\dots,M\}}\delta_{i,j,k}\leq 1, \, \forall i,k$. 
	This condition can be satisfied by having subsystems to back off from all other channels when they are granted access to one channel. Similar to what has been done with timers in \cite{Farjam:2019a,Farjam:2021,Farjam:2019b}, following the approach in Proposition~\ref{prop:p1}, the solution can be obtained by setting $m_{i,j,k}= {\CoIL}_{i,k}q_{i,j}$.
\end{remark}

%
%
%
%
\section{Scenario 2: Smart sensors}\label{sec:voi}
In this section, we consider the case of smart sensors which contain an embedded microprocessor. This allows them to do local computations and preprocess the raw measurements before transmission. We first discuss how the computational resources can be utilized to enhanced the estimation performance. Next, for the adopted triggering mechanism, the corresponding priority measure is derived and shown to be compatible with TOD. 

\subsection{Sensing and estimation}
It is well-known that preprocessing the raw measurements before transmission can enhance the estimation quality at the receiver side \cite{Gupta:2007, Schenato:2008}. Let $\mathcal{Y}_{i,k}=\{y_{i,0},\ldots,y_{i,k}\}$ denote the measurement history available to $\mathcal{S}_i$ at time $k$. The data packet sent by the sensor contains the MMSE state estimate, denoted by $\hat{x}_{i,k|k}^s$, which is calculated by running the following standard Kalman filter
\begin{subequations}
	\begin{align}
		\hat{x}_{i,k|k-1}^s &= A_{i}\hat{x}_{i,k-1|k-1}^s+B_iu_{i,k-1},\\
		P_{i,{k|k-1}}^s &= h_i(P_{i,{k-1|k-1}}^s)\simeq h_i(\overbar{P}_i),  \\
		K_{i,k}^s &= P_{i,{k|k-1}}^s C_i^T( C_iP_{i,{k|k-1}}^s C_i^T + V_i )^{-1}\simeq K_i^s, \label{eq:Kgain}\\
		\hat{x}_{i,k|k}^s &= \hat{x}_{i,k|k-1}^s + K_{i,k} (y_{i,k}-C_i\hat{x}_{i,k|k-1}^s),  \\
		P_{i,{k|k}}^s &= g_i\circ h_i(P_{i,{k-1|k-1}}^s)\simeq \overbar{P}_i, \label{eq:PposStv}
	\end{align}
\end{subequations}
By assuming that the pairs $(A_i,C_i)$ and $(A_i,W_i^{1/2})$ are observable and controllable, respectively, the \emph{a posteriori} error covariance in \eqref{eq:PposStv} converges exponentially fast to the unique positive semi-definite solution of $g_i\circ h_i (X)=X$ \cite{BrianD.O.Anderson:2012}. Let $\overbar{P}_i$ denote this solution and assume that the filter has already entered steady state and thus \eqref{eq:PposStv} can be written as $P_{i,{k|k}}^s = \overbar{P}_i$. As a result, the filter's gain in \eqref{eq:Kgain} is also time-invariant and we drop the time subscript and refer to it as $K_{i}^s$ in the subsequence.

For this scenario, the information available to the estimator is given by $\mathcal{I}_{i,k}=\{\delta_{i,0},\gamma_{i,0}\hat{x}_{i,0|0}^s ,\dots,\delta_{i,k},\gamma_{i,k}\hat{x}_{i,k|k}^s \}$ and the state estimate and error covariance are obtained by
\begin{align*}
	\hat{x}_{i,k|k}&= (A_i+B_iL_{i,\infty})^{t_{i,k}}\hat{x}_{i,k-t_{i,k}|k-t_{i,k}}^s,\\
	P_{i,k|k}&= h_i^{t_{i,k}}(\overbar{P}_i),
\end{align*}
where $t_{i,k}$ denotes the time elapsed since the last successful packet arrival, i.e.,
\begin{align}
	t_{i,k}\triangleq\min\{\kappa\geq0:\gamma_{i,{k-\kappa}}=1\}.
\end{align}
In essence, the estimator uses $\hat{x}_{i,k|k}^s$ if the packet arrives at $k$; or simply runs a prediction step otherwise.

\subsection{Priority measure and contention resolution}
Using the real-time data for deterministic or stochastic triggering of the events is more beneficial since it conveys additional information about the state of the system compared to VBT. However, this results in more energy consumption, since the sensors are required to constantly monitor the process and process the measurements. Herein, we show how the availability of real-time data and computational resources at the sensors can be utilized for solving Problem~\ref{problem:p1} in a distributed manner. The decision makers in this scenario are the smart sensors and the information available at each sensor $i$ can be described by ${\mathcal{I}}_{i,k}^s=\mathcal{Y}_{i,k}\cup t_{i,k}$ and thus $\mathcal{I}^k=\cup_{i\in\{1,\dots,N\}}{\mathcal{I}}_{i,k}^s$. Note that $t_{i,k}$ is known from ACK/NACK and allows the sensor to infer the state estimate and thus the applied inputs at the receiver side.
\begin{proposition} \label{prop:p2}
	Let $\check{e}_{i,k|k}\triangleq\hat{x}_{i,k|k}^s-(A_i+B_iL_{i,\infty})\hat{x}_{i,k-1|k-1}$ be the discrepancy between sensor's a posteriori state estimate and estimator's estimate in case it receives no data packet at $k$. Define VoI for subsystem $i$ at $k$ as
	\begin{align} \label{eq:VoI}
		{\VoI}_{i,k}\triangleq\tr\left(\Gamma_{i,\infty}\check{e}_{i,k}\check{e}_{i,k}^T\right).
	\end{align}
	Then, Problem~\ref{problem:p1} is solved by letting $\delta_{i^*,k}=1$ where 
	\begin{align} \label{eq:istrVoI}
		i^*=\argmax_{i\in\mathcal{F}_{k}}  \quad {\VoI}_{i,k}q_i.
	\end{align}
\end{proposition}
\begin{proof}
	Using the new information available to the decision makers, \eqref{eq:istrJ} can be written as
	\begin{align} \label{eq:istrJvoi}
		i^*=\argmax_{i\in\mathcal{F}_{k}}  \quad &\E\{{J}_{i,k}|{\mathcal{I}}_{i,k}^s,\delta_{i,k}=0\} \notag\\
		&\quad -\E\{{J}_{i,k}|{\mathcal{I}}_{i,k}^s,\delta_{i,k}=1\}.
	\end{align}	
Let $e_{i,k|k}^s\triangleq x_{i,k}-\hat{x}_{i,k|k}^s$ and thus the first term on the right hand side of \eqref{eq:istrJvoi} can be written as
\begin{align} \label{eq:Jik0voi}
			&\E\{{J}_{i,k}|{\mathcal{I}}_{i,k}^s,\delta_{i,k}=0\} -\tr\left(\Pi_{i,\infty} W_{i} \right) \notag \\ 
			&= \tr(\Gamma_{i,\infty}\E\{e_{i,k|k}e_{i,k|k}^T|{\mathcal{I}}_{i,k}^s,\delta_{i,k}=0\})\notag \\ 
			&= \tr(\Gamma_{i,\infty}\E\{(e_{i,k|k}^s+\check{e}_{i,k|k})(e_{i,k|k}^s+\check{e}_{i,k|k})^T|{\mathcal{I}}_{i,k}^s,\delta_{i,k}=0\} \notag\\
			&= \tr(\Gamma_{i,\infty}\E\{e_{i,k|k}^s{e_{i,k|k}^s}^T|{\mathcal{I}}_{i,k}^s,\delta_{i,k}{=}0\}){+}\tr(\Gamma_{i,\infty}\check{e}_{i,k|k}{\check{e}_{i,k|k}}^T) \notag \\
			&= \tr(\Gamma_{i,\infty}\overbar{P}_i)+\tr(\Gamma_{i,\infty}\check{e}_{i,k|k}{\check{e}_{i,k|k}}^T),
\end{align}
where the first equality is obtained by rearranging the terms in \eqref{eq:Jik} and the law of total expectation and the second equality follows from $e_{i,k|k}=e_{i,k|k}^s+\check{e}_{i,k|k}$. The facts that $\E\{e_{i,k|k}^s|{\mathcal{I}}_{i,k}^s\}=0$ and $\check{e}_{i,k|k}$ is deterministically given by
\begin{align} \label{eq:hate}
	\check{e}_{i,k|k}=\sum_{\tau=k-t_{i,k}}^k A_i^{k-\tau}K_i^s(y_{i,k}-\hat{x}_{\tau|\tau-1}^s),
\end{align}
where $K_i^s$ is the Kalman gain in \eqref{eq:Kgain}, yield the third equality. Using a similar approach for the second term on the right hand side of \eqref{eq:istrJvoi} yields
\begin{align} \label{eq:voi1q}
	&\E\{{J}_{i,k}|{\mathcal{I}}_{i,k}^s,\delta_{i,k}=1\} \notag \\ 
	&=\tr\left(\Pi_{i,\infty} W_{i} \right)+\tr(\Gamma_{i,\infty}\overbar{P}_i)q_i  \\
	&\qquad +(1-q_i)\left(\tr(\Gamma_{i,\infty}\overbar{P}_i)+\tr(\Gamma_{i,\infty}\check{e}_{i,k|k}{\check{e}_{i,k|k}}^T)\right),\notag
\end{align}
Rearranging the terms in \eqref{eq:Jik0voi} and substituting that and \eqref{eq:voi1q} in \eqref{eq:istrJvoi} completes the proof.
\end{proof}
Sensors are able to compute $\check{e}_{i,k|k}$ locally due to the decoupled dynamics. Therefore, ${\VoI}_{i,k}$ can also be calculated without requiring additional information from other subsystems. As a result, TOD can be implemented for solving Problem~\ref{problem:p1} in this setting as well by utilizing $m_{i,k}= {\VoI}_{i,k}q_i$ in \eqref{eq:TOD}. As aforementioned, optimality of the solution obtained by using this setup is not guaranteed. However, $f(\cdot)$ in \eqref{eq:TOD} can be fine-tuned for near-optimal performance as will be shown in Section~\ref{sec:results}.

\begin{conjecture} \label{prop:p3}
The system is Lyapunov mean-square stable when using the TOD scheme with $m_{i,k}= {\VoI}_{i,k}q_i$ if
		\begin{align} \label{eq:stblty}
			\lim\limits_{t\to\infty} \mu_i(t)^{1/t} <\frac{1}{{\sigma_{\max}^2(A_i)}},\quad \forall i\in\{1,\dots,N\},
	\end{align}
where $\mu_i(t)\triangleq \PR\{t_{i,k}=t\}$.
\end{conjecture}

\begin{proof}[Sketch of the proof/intuition]
	It is proved in \cite[Theorem 1]{Farjam:2021} that condition \eqref{eq:stblty} guarantees boundedness of $\E\{P_{i,k|k}\}$ which is sufficient for Lyapunov mean-square stability of the system. This condition was verified in a time-triggered system with smart sensors and a variance-based priority measure for contention resolution which is given by
	\begin{align} \label{eq:coilbar}
		\overbar{\CoIL}_{i,k}=q_i\tr\left(\Gamma_{i,\infty}\left[h_i^{t_{i,k-1}+1}(\overbar{P}_i)-\overbar{P}_i\right]\right).
	\end{align}	
	Since $\overbar{\CoIL}_{i,k}$ is independent of the real-time measurements, its value only depends on the number of consecutive packet dropouts which results in deterministic channel access decisions. This allows for modeling the packet arrival sequence as an ergodic Markov chain whose unique stationary distribution can be utilized to determine $\mu_i(t)$ and verify \eqref{eq:stblty}. The same method can be applied for a system with smart sensors using VBT and TOD with $m_{i,k}=\overbar{\CoIL}_{i,k}$, since the channel access decisions become deterministic. Although not analytically proved here, as intuitively expected, and confirmed by the simulation results, the use of real-time data for decision making, i.e., using $m_{i,k}= {\VoI}_{i,k}q_i$, improves performance. Hence, $\E\{P_{i,k|k}\}$ in this scenario is upper bounded by the one using $m_{i,k}=\overbar{\CoIL}_{i,k}$ which concludes the proof.
\end{proof}

%
%
%
%
\section{Numerical Results}\label{sec:results}

\subsection{The effect of priority measure on channel access decisions} \label{sub:resA}
To demonstrate the effect of the adopted priority measure on the channel access decisions, we consider an illustrative scenario where a wireless channel is shared between two subsystems with system matrices
\begin{align*}
	A_1 = \begin{bmatrix}
		1.1 & 0 \\
		0 & 0.9 \\
	\end{bmatrix}
	A_{2} = \begin{bmatrix}
		0.9 & 0 \\
		0 & 0.9 \\
	\end{bmatrix},
\end{align*}
and $B_i = C_i = Q_i = I_{2}$, $V_i=R_i=0.01I_{2}$, and $W_i=0.1I_{2}$ for $i\in\{1,2\}$. Moreover, the probability of successful transmission is assumed to be $q_1=0.85$ and $q_2=0.5$. We consider the case where an infinitesimal triggering threshold is implemented for both subsystems. In other words, both subsystems constantly compete for transmitting their data packet. This allows us to isolate the impact of the sensors' capabilities on the choice of priority measure $m_{i,k}$ in \eqref{eq:TOD} and thus on the contention resolution outcome. The contention is resolved by using 29-bit identifiers as it is the standard in CAN2.0B, where $n=20$ most significant bits represent ${\ID}_{i,k}^d$ while the remaining 9 bits allow the network to accommodate $2^9-1=511$ with unique static identifiers. 

\begin{figure}[h]
	\centering
	\includegraphics[width=.95\columnwidth]{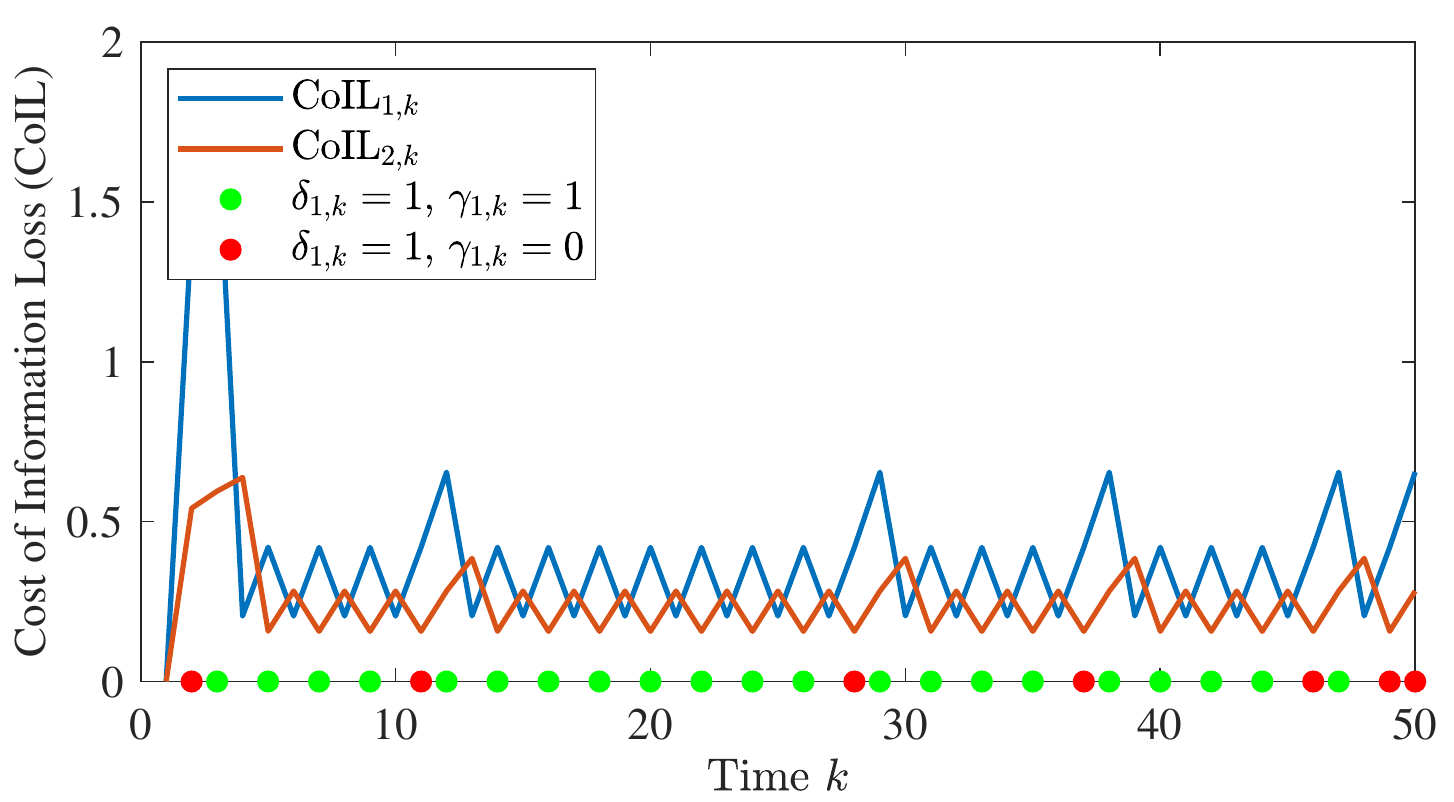}
	\vspace{-0.25cm}
	\caption{Impact of using $m_{i,k}={\CoIL}_{i,k}q_i$ in \eqref{eq:TOD} when two subsystems with conventional sensors constantly compete for transmission over a single wireless channel.}
	\label{fig:coil}
\end{figure}

\begin{figure}[h]
	\centering
	\includegraphics[width=.95\columnwidth]{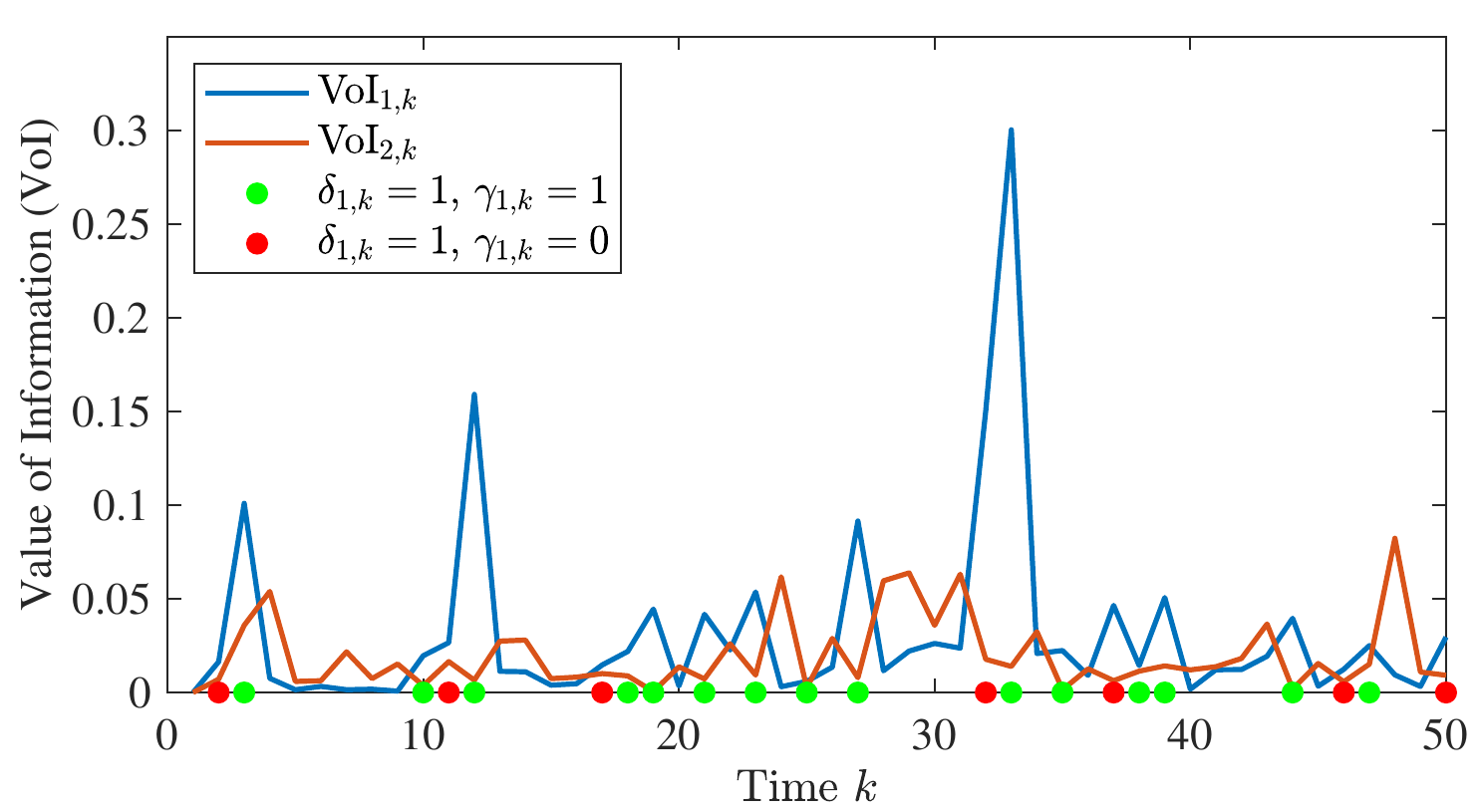}
	\vspace{-0.25cm}
	\caption{Two subsystems with smart sensors constantly compete for transmission over a single wireless channel based on $m_{i,k}={\VoI}_{i,k}q_i$ in \eqref{eq:TOD}.}
	\label{fig:voi}
\end{figure}

Fig~\ref{fig:coil} depicts the scenario of conventional sensors described in Section~\ref{sec:coil} and how utilizing $m_{i,k}={\CoIL}_{i,k}q_i$ affects the evolution of CoIL. For the setup considered here, we choose a simple function $f(m_{i,k})=\alpha m_{i,k}$ in \eqref{eq:TOD} with $\alpha=1000$ which results in unique dynamic identifiers and thus optimal channel access. As expected, Subsystem 2 wins the contention (shown as green and red dots) more frequently due to its unstable dynamics and larger growth rate of its CoIL. Furthermore, as CoIL is a function of the statistics of the random variables and unaffected by their real-time realization, it follows regular patterns. In contrast, as depicted in \ref{fig:voi} for the case of smart sensors, VoI is highly affected by the real-time realization of the random variables which is captured by the output $y_{i,k}$ in \eqref{eq:hate}. In addition to the irregular patterns, Subsystem 2 is granted channel access more frequently despite its stable dynamics. The impact of using both schemes on the long-term performance in terms of the average quadratic cost in \eqref{eq:Jinfty} is discussed next. 

\subsection{Computation capabilities and control performance}
\begin{figure}[h]
	\centering
	\includegraphics[width=.95\columnwidth]{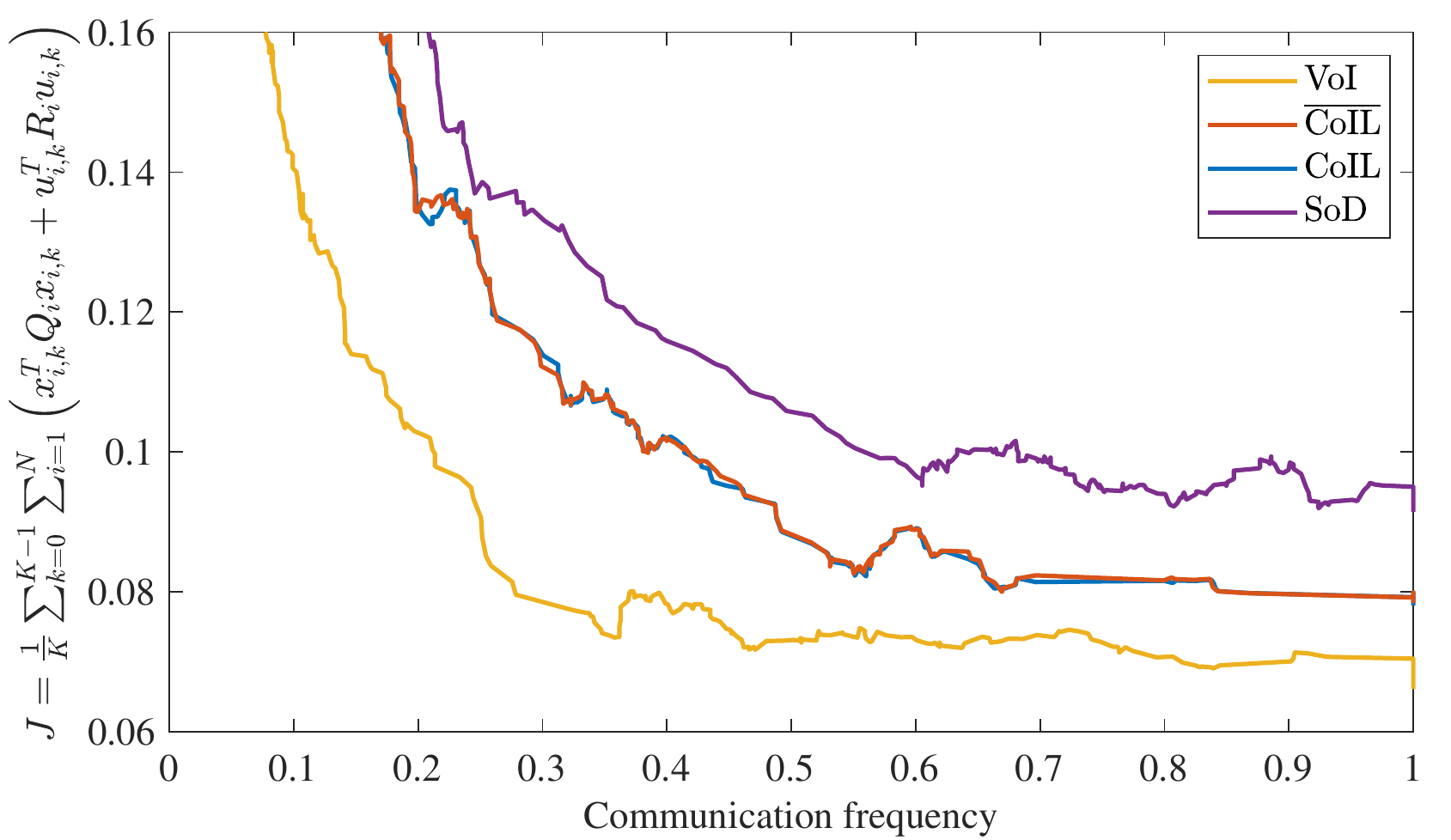}
	\vspace{-0.25cm}
	\caption{Average communication rate vs. quadratic cost for conventional and smart sensors.}
	\label{fig:res3}
\end{figure}

Herein, we demonstrate how the choice of sensors, which indicates the possible choices for the triggering mechanism and the priority measure for contention resolution, impacts control performance. To this end, average number of communication attempts, i.e., crossing the triggering threshold, versus the average quadratic cost in \eqref{eq:Jinfty} is depicted in Fig.~\ref{fig:res3}. The results are averaged over $1000$ simulations obtained for the system described in Section~\ref{sub:resA} on horizon $K=1000$.

Although all setups can be implemented with smart sensors, only two of them are compatible with conventional sensors, namely CoIL (Section~\ref{sec:coil}) and SoD. In SoD, transmission is triggered when the difference between the measured output and the last last successfully received one exceeds a threshold, i.e., $\norm{y_{i,k}-y_{i,k-t_{i,k}}}\geq \Delta$ and contention is based on $m_{i,k}=\norm{y_{i,k}-y_{i,k-t_{i,k}}}$. The additionally considered setups, which can only be realized by employing smart sensors, are denoted by VoI (Section~\ref{sec:voi}) and $\CoIL\overbar{P}$. The later refers to a scenario in which sensors preprocess their measurements similar to VoI, while the priority measure for contention resolution is variance-based, i.e., $m_{i,k}={\CoIL}_{i,k}q_i$ in \eqref{eq:TOD}.

As the results depicted in Fig.~\ref{fig:res3}, as the communication rate decreases, i.e., higher triggering threshold, performance deteriorates in all setups. Nevertheless, best performance is achieved when real-time processed measurements are utilized for event triggering and channel access as in VoI. Using real-time raw measurements, however, results in worst performance as indicated by SoD. Interestingly, when contention resolution is variance-based as in $\CoIL$ and $\overbar{\CoIL}$, transmission of the \emph{a posteriori} state estimate instead of the raw measurement by the sensor does does not lead to significant gain in terms of the average quadratic cost. Moreover, the effect of using smart sensors when considering scalar sensors communicating over a perfect channel is more significant in terms of the estimation error as shown in \cite{Trimpe:2015}. However, the improvements are less significant in terms of the average quadratic cost when the effect of packet dropouts and channel access decisions are explicitly taken into account.     

%
%
%
%
\section{Conclusions and Future Directions}\label{sec:conclusions}
We investigated distributed channel triggering mechanisms for WNCSs for conventional and smart sensors. First, we brought together several approaches proposed in the literature, drew the connection between them and extended their operation for vector systems in unreliable wireless multi-channel environments. More specifically, we proposed \emph{i)} the event-triggered version for CoIL for both conventional and smart sensors and \emph{ii)} the event-triggered version for VoI which allows for distributed implementation and takes into account the channel conditions. For all scenarios, in order to be able to consider the coupling arising due to the shared communication resources, we proposed prioritization schemes compatible with TOD protocol. The results showed that when smart sensors are used, utilizing measurement-based triggering and VoI as the priority measure leads to the best performance. However, when no computational power is available at the sensor side, as in case of conventional sensors, VBT and CoIL for prioritizing channel access offers the best performance. 

Future work will focus on event-triggered mechanisms for battery-operated transceivers in which the aim will be to further limit the energy consumption due to data transmission. Additional, we would like to consider transceivers with energy harvesting capabilities, in which the transmissions will be a function of the state of the battery levels. 

%
%
%
%

%
%
%
%
\bibliographystyle{IEEEtran}
\bibliography{refs}

\balance

%
%
%
%
\end{document}